\documentclass[prl,twocolumn,,aps,showpacs]{revtex4} 

\usepackage{amsmath}
\usepackage{graphicx}
\usepackage{dcolumn}
\usepackage{bm}

\newcommand{\cP}{{\cal P}}
\newcommand{\erf}{\mbox{erf}}
\newcommand{\glog}{\Lambda}
\newcommand{\gexp}{{\cal E}}

\begin{document}

\title{Maximum entropy approach to central limit distributions of correlated variables}{CLT\&MEP}

\author{Stefan Thurner$^{1,2}$ and Rudolf Hanel$^{1}$}

\affiliation{
$^1$ Complex Systems Research Group; HNO; Medical University of Vienna; 
W\"ahringer G\"urtel 18-20; A-1090; Austria \\
$^2$ Santa Fe Institute; 1399 Hyde Park Road; Santa Fe; NM 87501; USA\\
}

\date{Version \today}

\begin{abstract}
Hilhorst and Schehr recently presented a straight forward computation of limit distributions 
of sufficiently correlated random numbers \cite{hilhorst}. Here we present the analytical 
form of entropy which --under the maximum entropy principle (with ordinary constraints)-- provides these limit distributions.
These distributions are not $q$-Gaussians and can not be obtained with Tsallis entropy. 
\end{abstract}


\maketitle

\section{Introduction}
Classical statistical mechanics is tightly related to the central limit theorem (CLT).
For example if one interprets the velocity of an ideal-gas particle as 
the result of $N$ random  collisions with other particles, the velocity distribution of 
particles corresponds to a $N$-fold convolution some distribution of momenta exchanges. 
For any such distribution,  as long as it is centered, stable and has a second moment, 
the central limit theorem immediately guarantees Maxwell-Boltzmann distributions  
for  $N\to \infty$. 
Alternative to this  mathematical approach the same distribution can be 
derived from a {\em physical} principle, where Boltzmanns $H$-function (entropy)
gets maximized under the constraint that average kinetic energy is proportional 
to temperature, $k_BT$. This principle is referred to as the maximum entropy principle
(MEP). 
These results are trivial for the ideal-gas, or equivalently, for independent random numbers. 
However,  as soon as correlations come into play things become more  
involved on both sides: CLTs for correlated random numbers have  
regained strong interest \cite{moyano06,umarov06,baldovin07}. 
Recently a general and transparent method of obtaining limit distributions of correlated 
random numbers was reported for a wide class of processes \cite{hilhorst}. 
In the context of the MEP there basically exist three methods to arrive at 
non-Boltzmann (non-Gaussian) distributions (with reasonably many constraints). The first 
is associated with Tsallis entropy which produces  $q$-exponentials or $q$-Gaussians under 
entropy maximization \cite{tsallis88}, the second is based on the so-called $\kappa$-logarithm 
\cite{kaniadakis} leading to a three-parameter class of distribution functions. 
Recently a method was introduced to constructively design entropy functionals 
which -- under maximization under ordinary constraints --  produce {\em any } 
plausible distribution function \cite{hanel07}. This was later shown to be a thermodynamically 
relevant entropy \cite{abe08}. This entropy is a generalization of previous generalizations 
of Boltzmann-Gibbs entropies therefore we call it {\em generalized-generalized} entropy, $S_{gg}$. 

In this paper we show that CLTs and the MEP can be brought into 
a consistent framework for correlated variables.
We start by reviewing the CLT for sums of correlated random numbers, following 
\cite{hilhorst}, and the MEP derivation for 
arbitrary distribution functions following, \cite{hanel07}. We then give the explicit form of the 
entropy leading to Hilhorst and Schehr's limit distributions.

\subsection{Limit theorems of correlated random numbers}\label{sechillimthe}

The idea in \cite{hilhorst} is to consider a totally symmetric correlated Gaussian $N$-point process 
\begin{equation}
P_N({\bf z})=\frac{e^{-\frac{1}{2}\left(z,M^{-1} z\right)}}{\sqrt{(2\pi)^N\det(M)}}
\quad,
\label{NGauss}
\end{equation}
with ${\bf z}=(z_1,\dots,z_N)$ and $M$ the covariance matrix. 
This stochastic process is used as a reference process for some other totally symmetric $N$-point 
distribution $\tilde P_N(u)$ which is related to $P_N(z)$ 
by a simple transformation of variables $u_i=h(z_i)$, for all $i=1,\dots,N$.
Total symmetry dictates the form the covariance matrix 
\begin{equation}
M_{ij}=\delta_{ij}+\rho (1-\delta_{ij})\quad,
\label{covmat}
\end{equation}
for $\rho \in \left(0,\,1\right]$
with the inverse, $M^{-1}_{ij}=a \delta_{ij}- b (1-\delta_{ij}) $, 
where 
$a=\frac{1+(N-2)\rho}{(1-\rho)(1+(N-1)\rho)}$, and 
$b=\frac{\rho}{(1-\rho)(1+(N-1)\rho)}$.
Straight forward calculation yields that the marginal probability
is a Gaussian with unit variance, 
 \begin{equation}
    P_1^{\mbox{Gauss}}(z_1)=\int dz_2\dots dz_N\,\, P_N(z_1,z_2,\dots,z_N)
    \quad .
    \label{unitgauss}
 \end{equation}
This allows to construct the set of variables $u_i$ from $z_i$
via the transformation of variables
 \begin{equation}
   \int_{0}^{u_i} du'\,\, P_1(u')=\int_{0}^{z_i} dz'\,\, P_1^{\mbox{Gauss}}(z')\quad,
   \label{hilvartrans}
 \end{equation}
$P_1$ being the one-point distribution of the $u$ variables.
Consequently, there is 
a unique function $h$, such that 
 \begin{equation}
    u_i=h(z_i) \quad.
 \end{equation}
The distribution of the average of the $u$ variables,  
\begin{equation}
\bar u=\frac{1}{N}\sum_{i=1}^N u_i\quad,
\label{hilavgvar}
\end{equation}
is thus found in terms of an integration over all $z_i$
 \begin{equation}
   \cP(\bar u)=\int d{\bf z}\,\, P_N({\bf z}) \,\, \delta\left(\bar u-\frac{1}{N}\sum_{i=1}^N h(z_i) \right)\quad,
   \label{hilavgdist}
 \end{equation}
where $d{\bf z}=dz_1\dots dz_N$.
After some calculation one arrives at the general result \cite{hilhorst}, 
 \begin{equation}
  \cP(\bar u)=  \left( \frac{1-\rho }{2 \pi \rho } \right)^{\frac12}   |Êk' (\nu_*(\bar u))|^{-1} \exp \left( - \frac{1-\rho }{2 \rho } [\nu_*(\bar u) ]^2 \right) 
  ,
  \label{hilavgdist2}
 \end{equation}
where $\nu_*$ is defined as the zero of the function  
 \begin{equation}
   k(\nu)= \bar u -\frac{1}{\sqrt{2\pi } }\int dw \, e^{- \frac{w^2}{2}   } h\left(  (w+\nu)\sqrt{1-\rho } \right)  
  \quad, 
  \label{hilk}
 \end{equation}
and $k'(x)=d/dx \,\,  k(x)$.
For symmetric one-point distributions $P_1$, $h$ and $\nu_*$ are both 
antisymmetric. Moreover it is seen that
$   \nu_*'(\bar u)=-k'(\nu_*(\bar u))^{-1}\geq 0  $,
such that 
 \begin{equation}
  \cP(\bar u)= \frac12\frac{d}{d\bar u} \erf\left(\sqrt{\frac{1-\rho}{2\rho}} \nu_*(\bar u) \right) \quad.
  \label{hilavgdist2x1}
 \end{equation}

\subsection{Generalized-generalized entropy}

The presently most general form of entropy that is consistent with the maximum entropy condition 
reads in dimensionless notation \cite{hanel07},
  \begin{equation}
     S_{gg}[P] = - \sum_i P(z_i) \Lambda (P(z_i)) + \eta[P] 
     \label{ent1}
  \end{equation}
with
  \begin{equation}
     \eta[P]=  \sum_i \int_0^{P(z_i)} dx \, x \frac{d \Lambda(x) }{dx} + c \quad ,
     \label{eta}
  \end{equation}
which can be rewritten as
  \begin{equation}
      S_{gg}[P] = - \sum_i   \int_0 ^{P(z_i)} dx\,\, \Lambda (x) +c \quad . 
     \label{ent2}
  \end{equation}
In these equations, $P(z_i)$ is a normalized 
distribution function of some parameter set $z$ 
\footnote{For continuous variables $z$ replace  $\sum_i\to\int dz$.}
In physics this could be e.g. energy or velocity. 
$\Lambda(x)$ is integrable in each interval $[0,P(z_i)]$. It can be
seen as a Ògeneralized logarithmÓ satisfying $\Lambda(x)<0$ and
$d\Lambda(x)/dx>0$ for $0<x<1$,  and $x\Lambda(x)\to 0 $ ($x\to 0+$),
making $S_{gg}[P]$ non-negative and concave. 
$c$ is a constant, which ensures that $S_{gg}[P_0]=0$ for a completely ordered state, i.e.
$c=-\int_0^1dx \Lambda (x)$.

A generalized maximum entropy method, given the
existence of some arbitrary  stationary distribution function, $\tilde P(z_i)$, is
formulated as 
 \begin{equation}
    \frac{\delta G }{\delta P(z_i) }  \ | _{P=\tilde P } =0
 \end{equation}
with
 \begin{equation} 
G\equiv S_{gg}[P] -\alpha\left\{ \sum_i P(z_i)  -1 \right\} -\beta \left \{ \sum_i f(z_i)P(z_i)  -U \right \} 
    , 
	\label{fvarprinc}
 \end{equation}
where $\alpha$ and $\beta$ are Lagrange multipliers, and $U$ denotes the expectation of function $f$, which 
depending on the problem may be a particular moment of $z$.
The stationary solution to this problem is given by
  \begin{equation}
    \tilde  P(z_i) = \gexp \left( -\alpha -\beta f(z_i) \right) \quad ,
    \label{gexp}
  \end{equation}
where $\gexp(x)$ is a Ògeneralized exponentialÓ, which is
the inverse function of $\Lambda(x)$: $\gexp(\Lambda(x))=\Lambda(\gexp(x))=x$.  
In other words, $\Lambda(x)$ from Eq. (\ref{ent1}) is chosen as the inverse function
of the stationary (observed) distribution function.

The form of entropy in Eq. (\ref{ent2}) is enforced by the maximum entropy 
principle. Under  
variation the first term on the right hand side of Eq. (\ref{ent1})
yields $\frac{d}{dx} P \Lambda(P) = \Lambda+ P \frac{d \Lambda }{dx}$. The term $P \frac{d \Lambda }{dx}$ can 
neither get absorbed into the logarithmic terms, nor by the constants. As long as this 
term is present, the only solution to the MEP is $\Lambda=\ln$. The idea of the generalized-generalized 
entropy is to introduce $\eta$ (which modifies entropy), such that the term $P \frac{d \Lambda }{dx}$ cancels 
out exactly under the variation, for details see \cite{hanel07}. 
It has been shown explicitly that the first and second laws of thermodynamics are 
robust under this generalization of entropy  \cite{abe08}. 
For the discussion below note that introducing a scaling factor $\zeta$ in the argument, 
 \begin{equation}
   S_{gg}[P]=- \sum_i \int\limits_0^{P(z_i )}dx\,\,\glog(\zeta x) 
   \quad ,
   \label{g3ent}
 \end{equation}
corresponds to distributions of the form
 \begin{equation}
   P(z)=\frac{1}{\zeta}\gexp(-\alpha -\beta f(z) )  \quad.
   \label{g3dist}
 \end{equation}
$\zeta$ and $\alpha$ can be seen as two alternative normalization parameters.
For instance the condition $\zeta=1$ leads to a normalization of the distribution 
as discussed in \cite{hanel07, abe08}, while
the condition $\alpha=0$ leads to a normalization of $P$ where $\zeta$ plays the role of
the {\em partition function}. 

\begin{figure*}[htp]
\includegraphics[width=4.5cm]{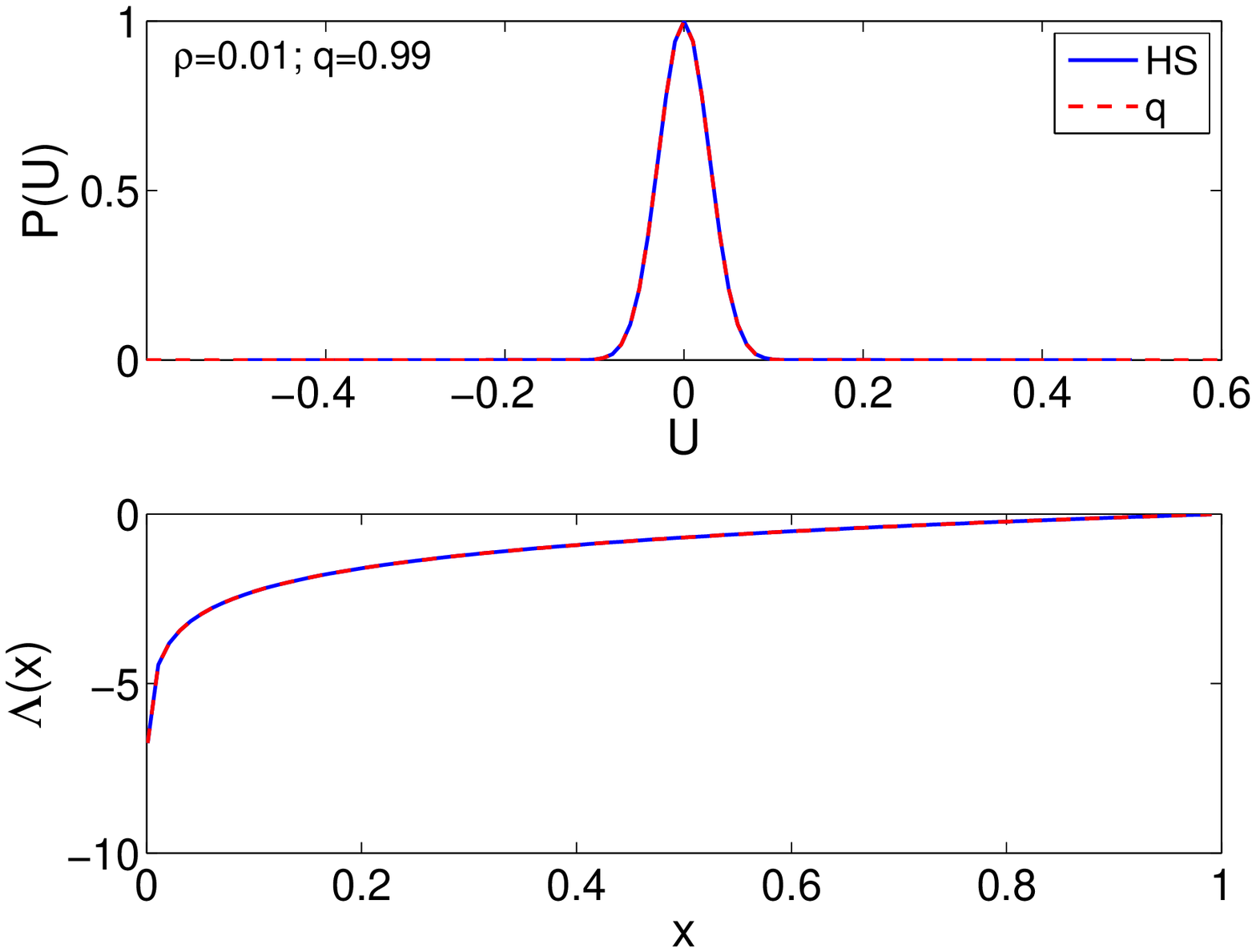}
\hspace{0.5cm}
\includegraphics[width=4.5cm]{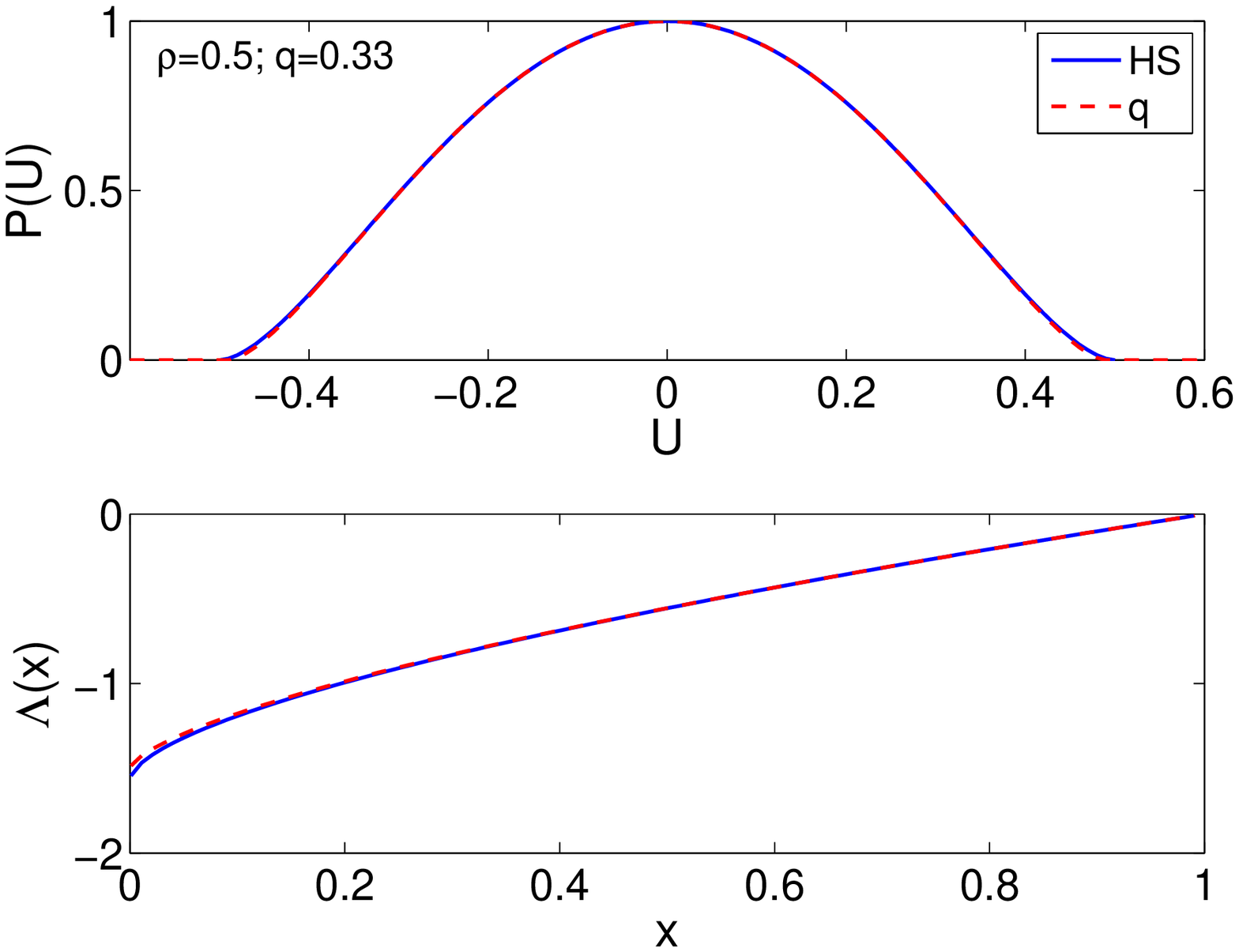}
\hspace{0.5cm}
\includegraphics[width=4.5cm]{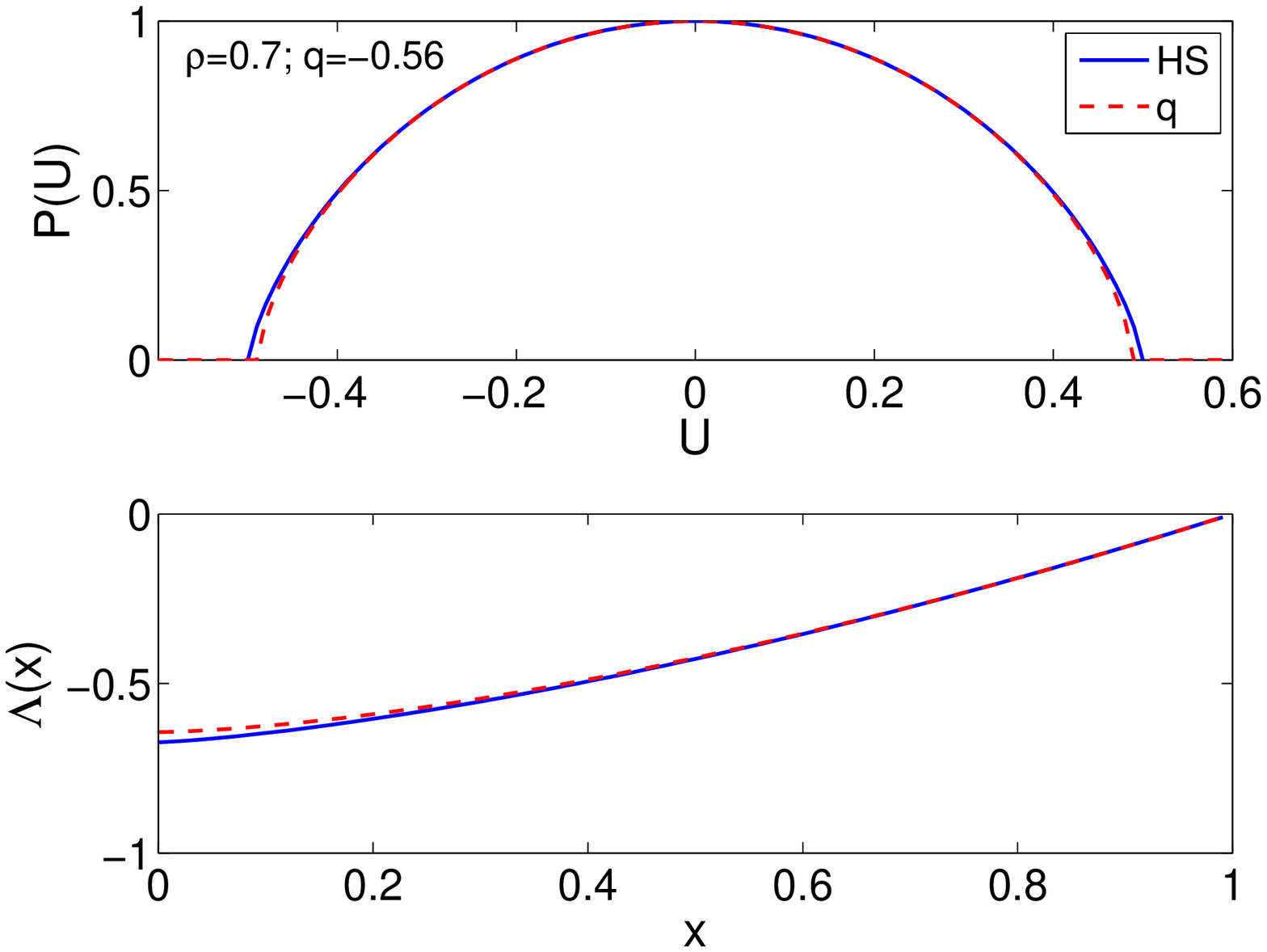}
\vspace{-0.3cm}
\caption{
Top: Limit distributions {\it al la} Hilhorst-Schehr for $\rho=0.001$ (left) $0.5$ (middle), and $0.7$ (right) (lines).
Broken lines are $q$-Gaussians with a $q$ from the reported best-fit value of 
$q=\frac{1-\frac53 \rho }{1-\rho }$, \cite{hilhorst}.   
Bottom: Generalized logarithms necessary to provide Hilhorst-Schehr distribution functions
under the maximum entropy principle (lines). Broken lines are $q$-logarithms 
$\ln_q(x)=(x^{1-q}-1)/(1-q)$ for  the same $q$ values.
}
\label{fig}      
\end{figure*}

\section{Entropy for limit distributions}

Let us now properly identify the distributions $P$ and $\cP$ from Eqs.  (\ref{hilavgdist2})  and (\ref{g3ent}). 
First, note that the limit distributions in \cite{hilhorst} are 
centered and symmetric thus the first moment does not provide any information. 
We therefore choose $f$ from Eq. (\ref{fvarprinc}) to be $f(z)=z^2$. 
Second, the one point distribution $P_1$ has fixed variance and so does $\cP(\bar u)$ 
in Eq. (\ref{hilavgdist2}), whereas distributions obtained through the
MEP Eq. (\ref{g3dist}) scale with a function of the ''inverse temperature'', $\beta$.
To take this into consideration for the identification of $P$ and $\cP$ we need  a simple 
scale transformation $\bar u=\lambda z$, where $\lambda(\beta)$ depends explicitly on $\beta$.
Consequently, $\cP(\bar u)\to \lambda \cP(\lambda z)$ and
 \begin{equation}
    \lambda \cP(\lambda z)=\frac1\zeta \gexp\left(-\alpha-\beta z^2\right)
   \quad .
   \label{leftrightid}
 \end{equation}
%
This particular identification and the independence of Lagrange parameters 
requires the normalization condition
$\alpha=0$ for the limit distribution Eq. (\ref{g3dist}).
Further, to determine $\zeta$ and $\lambda$ we use two conditions usually valid for generalized
exponential functions, $\gexp(0)=1$ and $\gexp'(0)=1$. This leads to 
 \begin{equation}
   \zeta^{-1}=   \lambda\left( \frac{1-\rho }{2 \pi \rho } \right)^\frac12 \nu_*'(0) 
   \quad {\rm and } \quad
    \lambda= \gamma \sqrt{\beta} 
   \quad , 
   \label{zeta}
 \end{equation}
with $ \gamma ^2\equiv \frac{2\rho\nu_*'(0)}{\nu_*'(0)^3(1-\rho)-\nu_*'''(0)\rho} $.
The generalized exponential can now be identified as
 \begin{equation}
  \gexp\left(x\right) =  \frac{\nu_*'(\gamma \sqrt{-x})}{\nu_*'(0)}\exp \left( - \frac{1-\rho }{2 \rho } 
  [\nu_*(\gamma \sqrt{-x}) ]^2 \right)
  \quad,
  \label{GEeq}
 \end{equation}
This uniquely defines $\gexp$ on the domain $\left(-\infty, 0\right]$.
Finally, the generalized logarithm $\glog$ is uniquely defined on
the domain $\left(0,1\right]$ as the inverse function of $\gexp$ and can be given explicitly for
specific examples.

\subsection{An example}

The special case  of a block function $P_1(u_j)=1$ for 
$-\frac12 \leq u_j \le  \frac12$ was discussed in \cite{hilhorst}. This choice implies 
 \begin{eqnarray} 
   k'(\nu) &=& -\frac{\kappa e^{-\kappa^2 \nu^2 } }{\sqrt{\pi }} \quad , \quad \kappa\equiv \left(  \frac{1-\rho }{2(2-\rho)}  \right)^{\frac12 } \nonumber \\
   k(\nu)  &=& \bar u -\frac12 \erf(\kappa \nu)  \quad , \quad      
    \nu_* = \kappa^{-1} \erf^{-1} (2\bar u)  , 
 \end{eqnarray}
and  the limit distribution Eq. (\ref{hilavgdist2}) becomes 
 \begin{equation}
   \cP(\bar u)=\left(  \frac{2-\rho }{\rho } \right)^{\frac12 } \exp \left( -\frac{2(1-\rho)}{\rho } [\erf^{-1} (2\bar u )]^2  \right)
   \quad .
   \label{hilavgdist3}
\end{equation}
This block function has been used earlier \cite{thristleton} where it was conjectured on numerical evidence that 
the limiting distribution would be a $q$-Gaussian. This is obviously ruled out by Eq. (\ref{hilavgdist3}), 
however, actual discrepancy is small, see Fig. \ref{fig}. 
For this example 
Eq. (\ref{GEeq}) becomes 
\begin{equation}
  \gexp(x)\equiv \exp \left(-(\pi \gamma ^2)^{-1}\left[\erf^{-1}(2\gamma \sqrt{-x})\right]^2 \right) 
  \quad ,
  \label{gexp1}
 \end{equation}
where $\gamma=\sqrt{\rho/(2\pi(1-\rho))}$. The associated generalized logarithm can be 
explicitly given on the domain $\left(0, 1\right]$ 
 \begin{equation}
   \glog(x)=-\left[ (2\gamma)^{-1} \erf \left( \gamma  \sqrt{ - \pi \ln x   }\right)  \right]^2
   \quad .
   \label{glog2}
 \end{equation}
It is compared to $q$-logarithms in Fig. \ref{fig}; the discrepancy is small but visible.  

\section{Discussion}

Historically it was hypothesized on numerical evidence \cite{thristleton} that 
limit distributions of sums of correlated random numbers generated along the lines of 
Eq. (\ref{unitgauss}) might be deeply related to $q$-Gaussians, thus lending 
fundamental support for $q$-entropy. 
In \cite{hilhorst} it was shown that this is not the case. 
If $q$-entropy does not  lead to the exact limit distributions under the MEP, 
which form of entropy does? 
Here we constructively answer this question by using a recently proposed generalization 
of $q$-generalized entropy \cite{hanel07}, which is thermodynamically relevant \cite{abe08}.   
Interestingly, a similar program has been carried out some time ago for Levy stable distributions, 
which result from a generalized CLT where higher 
momenta do not exist. There  it was shown that the corresponding entropy 
functional  is Tsallis entropy \cite{abe00}, however under $q$-constraints.

Supported by Austrian Science Fund FWF Projects P17621 and P19132.

\end{document}